      [1999/12/01 v1.4c Il Nuovo Cimento]
\documentclass{cimento}
\usepackage{mcite}
\usepackage{amsmath}

\newcommand{\be}{\begin{equation}}
\newcommand{\ee}{\end{equation}}
\newcommand{\bea}{\begin{eqnarray}}
\newcommand{\eea}{\end{eqnarray}}
\newcommand{\beq}{\begin{eqnarray}}
\newcommand{\eeq}{\end{eqnarray}}

\newcommand{\Op}{\mathcal{O}} 

\newcommand{\Dlr}{\buildrel \leftrightarrow \over D\raise-1pt\hbox{}}

\usepackage{graphicx}  

\title{Nucleon Structure using lattice QCD}
\author{C.~Alexandrou\from{ins:1}\from{ins:2}\thanks{Speaker}\ETC,
M.~Constantinou\from{ins:1},
V.~Drach\from{ins:3},
K.~Hatziyiannakou~\from{ins:1},
K.~Jansen\from{ins:3}\from{ins:1},
C.~Kallidonis\from{ins:1}\from{ins:2},
G.~Koutsou\from{ins:2},
T.~Leontiou\from{ins:4} \atque
A.~Vaquero\from{ins:2}
       }

\instlist{ 

  \inst{ins:1} Department of Physics, University of Cyprus, PO Box
  20537, 1678 Nicosia, Cyprus

  \inst{ins:2} Computational-based Science and Technology Research
  Center, The Cyprus Institute, P.O. Box 27456, CY-1645 Nicosia,
  Cyprus 
  
  \inst{ins:3} NIC, DESY, Platanenallee 6, D-15738 Zeuthen, Germany 
  \inst{ins:4} General Department, Frederick University, 1036 Nicosia, Cyprus
}
\begin{document}

\maketitle

\begin{abstract}
A review of recent nucleon structure calculations within lattice QCD
is presented. The nucleon excited states, the  axial charge, the isovector momentum
fraction and helicity distribution are discussed, assessing the
methods applied for their study, including approaches to evaluate the
disconnected contributions.  Results on the
spin carried by the quarks in the nucleon are also presented.
\end{abstract}

\section{Introduction}
During the last decade, results from simulations of QCD have emerged
that already provide essential input for a wide range of strong
interaction phenomena as, for example, the QCD phase diagram, the
structure of hadrons, nuclear forces and weak decays.  In this
presentation, we focus on hadron structure calculations using
state-of-the art lattice QCD
simulations~\cite{Alexandrou:2010cm, Alexandrou:2011iu, Alexandrou:2012da}.

Understanding nucleon structure from first principles is considered a
milestone of hadronic physics. A rich experimental program has been
devoted to its study, starting with the measurements of the
electromagnetic nucleon form factors initiated more than 50 years
ago. Reproducing these key observables within the lattice QCD
formulation is a prerequisite for obtaining reliable predictions on
observables that explore physics beyond the standard model.  In
particular we discuss three fundamental quantities, which can be
easily extracted from nucleon matrix element calculations on the
lattice, namely the nucleon axial charge, the isovector momentum
fraction and the helicity moment. The reason for this choice is that
they are known experimentally and they all can be directly determined
at momentum transfer squared $q^2=0$. Thus there is no ambiguity
associated with having to fit the $q^2$-dependence of the form factor
(FF), as for example, in the case of the anomalous magnetic moment
where one needs to fit the small $q^2$-dependence of the magnetic
FF. In addition, only the connected diagram, for which established
lattice methods exist for its calculation, contributes.

For the discussion of the spin content of the nucleon however, one
needs to calculate the contributions carried by the u- and d-
quarks. Contributions from fermion loops need to be taken into account
in this case. There are on-going efforts to compute these so called
disconnected contributions, which will be briefly discussed
here. However, in the final results shown in this work for the quark
spin these contributions will be neglected.
 
\section{Methodology}
It is convenient
to decompose the nucleon matrix elements, $\langle
N(p',s')|\Op_\Gamma^{\mu\mu_1\ldots\mu_n}|N(p,s)\rangle$, into generalized form factors
(GFFs) as follows \beq 
u_N(p',s') \Biggl[\sum_{i=0,2,\ldots}^{n}&\left(A_{n+1,i}(q^2)
  \gamma^{\{\mu}+{ B_{n+1,i}(q^2)} \frac{i\sigma^{\{\mu
      \alpha}q_\alpha} {2m} \right)q^{\mu_1}\ldots q^{\mu_{i}}
  \overline P^{\mu_{i+1}}\ldots\overline P^{\mu_n\}}\Biggr.  \nonumber
  \\ &+\Biggl.{\rm mod}(n,2) {C_{n+1,0}(q^2)} \frac{1}{m} q^{\{
    \mu}q^{\mu_1}\ldots q^{\mu_n\}} \Biggr] u_N(p,s).  \eeq For
$p^\prime=p$ one has the $n^{\rm th}$ moment of the unpolarized parton
distribution, $\langle x^n \rangle _q$. A similar expression can be
written for $\Op_{\Delta q}^{\mu\mu_1\ldots\mu_{n}}$ in terms of
$\tilde A_{ni}(q^2)$ and $\tilde B_{ni}(q^2)$. The ordinary nucleon
form factors are then special cases of GFFs given by \beq A_{10}(q^2)
&=& F_1(q^2)=\int_{-1}^1 dx H(x,\xi,q^2), \quad B_{10}(q^2) =
F_2(q^2)=\int_{-1}^1 dx E(x,\xi,q^2)\nonumber \\ \tilde A_{10}(q^2)
&=& G_A(q^2)=\int_{-1}^1 dx \tilde{H}(x,\xi,q^2), \quad \tilde
B_{10}(q^2) = G_p(q^2)=\int_{-1}^1 dx \tilde{E}(x,\xi,q^2) \eeq while
$A_{n0}(0)$, $\tilde A_{n0}(0)$, $A^T_{n0}(0)$ are moments of parton
distributions, e.g. $\langle x \rangle_q = A_{20}(0) $ and $\langle x
\rangle_{\Delta q} = \tilde A_{20}(0)$ are the first moments of the
spin independent and helicity distributions. With knowledge of these
moments, one can evaluate the quark contributions to the nucleon
spin using the decomposition $J_q = \frac{1}{2}[ A_{20}(0) +
  B_{20}(0)]=\frac{1}{2}\Delta \Sigma+ L_q$ and $\Delta
\Sigma=\tilde{A}_{10}$.  The nucleon spin sum rule $\frac{1}{2} =
\frac{1}{2}\Delta \Sigma+ L_q+J_g$ allows the determination of the
gluon contribution $J_g$.

\begin{figure}[h!]
    \begin{minipage}[t]{0.475\linewidth}
      \includegraphics[width=0.9\linewidth]{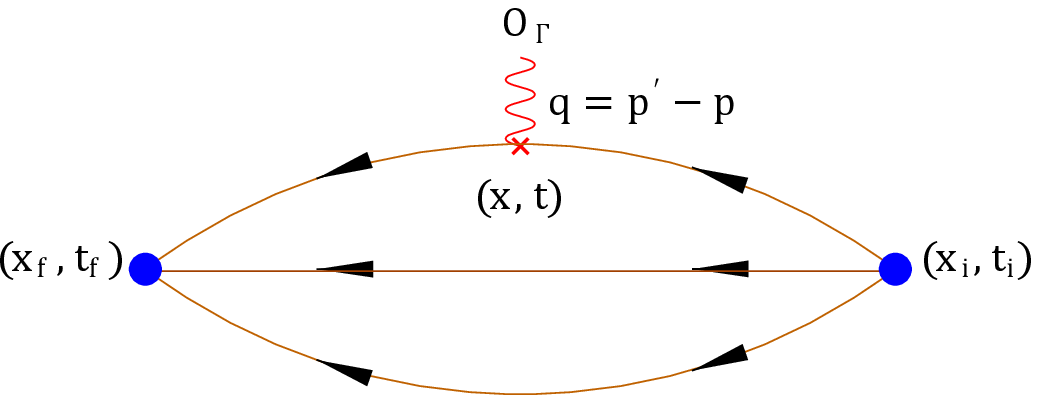}
      \caption{Connected contribution to a baryon 3-point
        function.}
      \label{fig:conn three_pt}
    \end{minipage}\hfill
    \begin{minipage}[t]{0.475\linewidth}
      \includegraphics[width=0.9\linewidth]{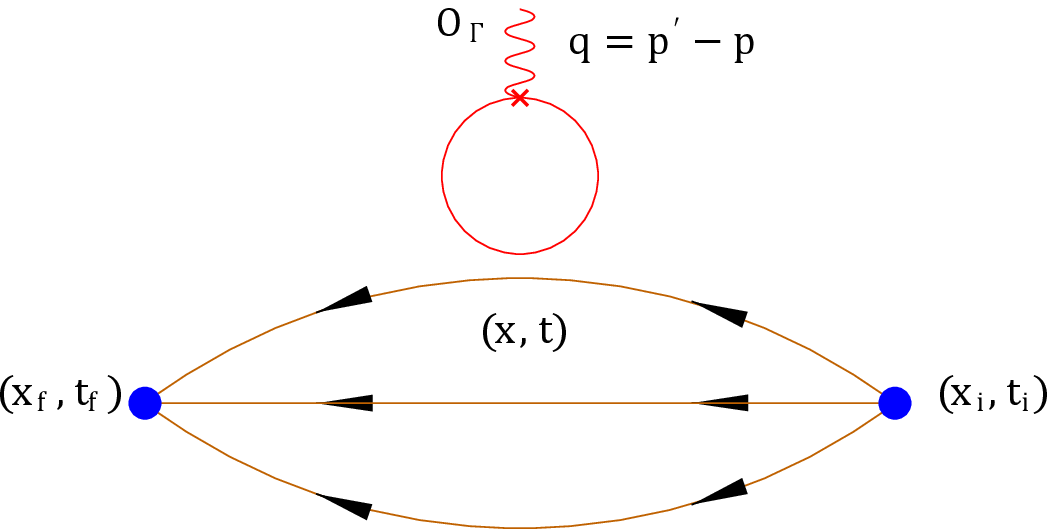}
      \caption{Disconnected contribution to a baryon 3-point
        function.}
      \label{fig:disconn three_pt}
    \end{minipage}
\end{figure}

In lattice QCD, one proceeds with the calculation of nucleon matrix
elements, by creating a state with the quantum numbers of the nucleon
$J_N^{\alpha}$ at an initial time, $t_i=0$ (source), inserting a current at some intermediate
time, $t$, and subsequently annihilating the nucleon at the final
time, $t_f$, (sink). Such a three-point function is shown in
Figs.~\ref{fig:conn three_pt} and \ref{fig:disconn three_pt}, where the lines represent fully dressed quark
propagators. For a general operator $\mathcal{O}_\Gamma$, one is
interested in the momentum-dependent matrix element:
\begin{equation}
  G^{\mu\mu_1...\mu_n}(\Gamma^\nu;t_f;t;\vec{q}) = \sum_{\vec{x}_f,\vec{x}_i}\Gamma^\nu_{\beta\alpha}\langle J^{\alpha}_{N}(\vec{x}_f, t_f)\mathcal{O}^{\mu\mu_1...\mu_n}(\vec{x}, t)\bar{J}^\beta_N(\vec{0}, 0)\rangle e^{-i\vec{q}\cdot\vec{x}}
\end{equation} 
where the projection matrices $\Gamma^\nu$ are $ \Gamma_0 =
\frac{1}{4}(1+\gamma_0),\quad\Gamma_k=i\Gamma_0\gamma_5\gamma_k,\,k=1,2,3.
$ Depending on the quark structure of the operator
$\mathcal{O}_\Gamma$, such a matrix element is a combination of
connected and disconnected contributions, as shown in
Figs.~\ref{fig:conn three_pt} and \ref{fig:disconn three_pt}
respectively.

\begin{figure}
  \vspace{-1ex}
    \begin{minipage}{0.475\linewidth}\vspace*{0.5cm}
      \includegraphics[width=1\linewidth]{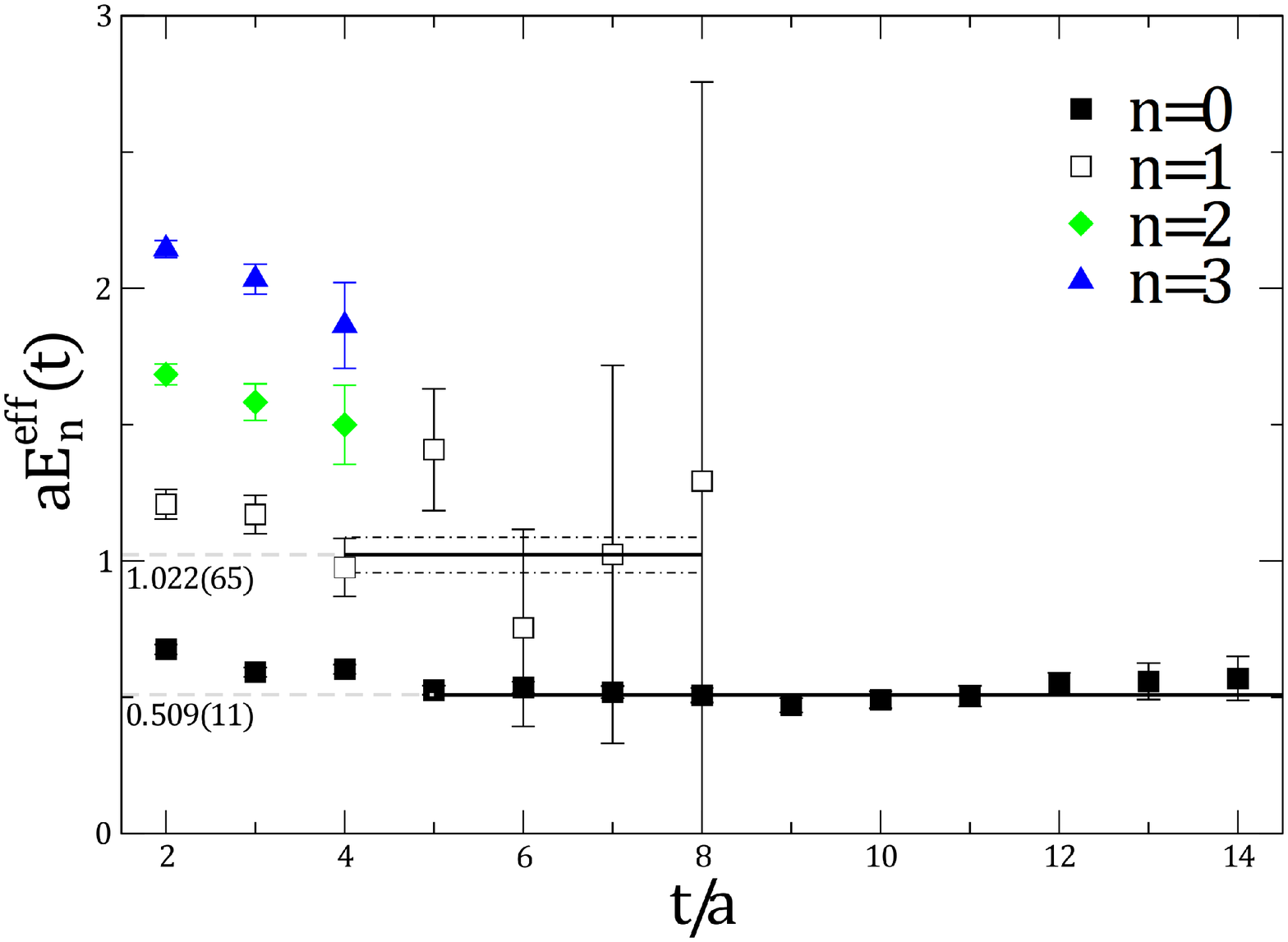}
      \caption{The four lowest states of the nucleon in the positive
        parity channel extracted using $N_f=2$ twisted mass fermions (TMF) with pion mass 308~MeV. The solid
        lines and bands show the value and error extracted from a fit
        in the plateau region.}
      \label{fig:spectrum}
    \end{minipage}\hfill
    \begin{minipage}{0.475\linewidth}
      \includegraphics[width=1\linewidth]{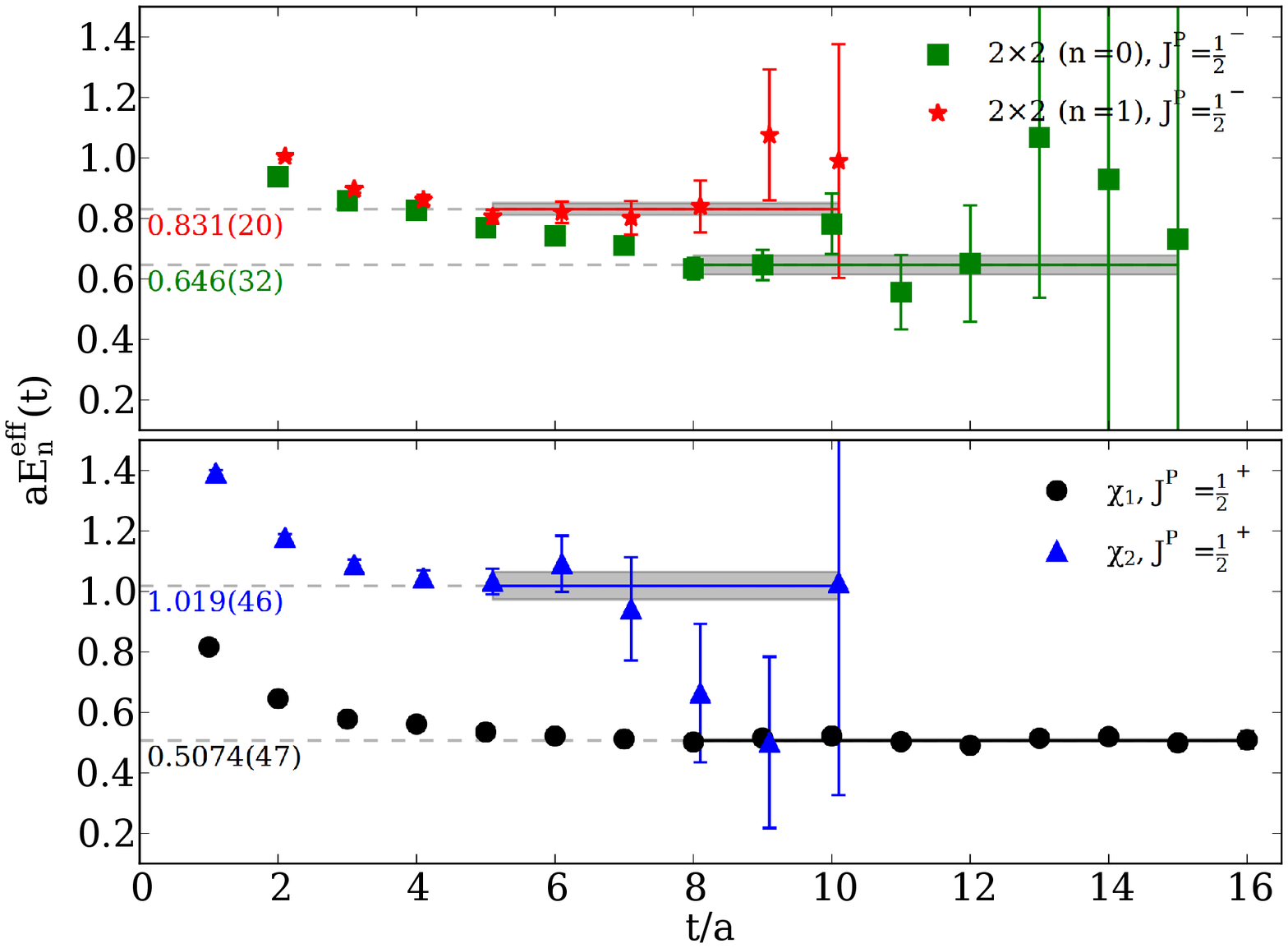}
      \caption{The two lowest states of the nucleon in the positive
        (lower) and negative (upper) parity channels using two dynamical
        flavors ($N_f=2$) of TMF gauge ensembles with pion mass
        308~MeV.}
      \label{fig:spectrum2}
    \end{minipage}
\end{figure}

The nucleon matrix element is obtained in the limit where the
time separation between the current insertion and
the source and the sink are large enough 
 such that only
the ground state matrix element dominates.
An appropriately defined ratio 
\bea
  R^{\mu}(\Gamma^\nu;t_f;t;\vec{q}) = \frac{G^{\mu}(\Gamma^\nu;t_f;t;\vec{q})}{G(\vec{0},t_f)}\sqrt{\frac{G(\vec{q},t_f-t)G(\vec{0},t)G(\vec{0},t_f)}{G(\vec{0},t_f-t)G(\vec{q},t)G(\vec{q},t_f)}}\xrightarrow[t_f-t\gg 1]{t\gg 1} \Pi^{\mu}(\Gamma^\nu;\vec{q})
\label{ratio}
\eea
thus becomes time independent (``plateau region'') and
unknown overlaps of the nucleon state with the trial states are canceled.
In Eq.~(\ref{ratio}) we have taken the final momentum $\vec{p}^\prime = 0 $. The two-point function appearing in Eq.~(\ref{ratio}) is given by: 
\be 
G(\vec{q},t_f) =
\sum_{\vec{x}_f}\Gamma_0^{\beta\alpha}\langle J^{\alpha}_{N}(\vec{x}_f,
t)\bar{J}^\beta_N(\vec{0}, 0)\rangle e^{-i\vec{q}\cdot\vec{x}_f}.
\label{two_pt}
\ee
 Energies of hadronic states are extracted from two-point functions
by taking the ratio $E_{\rm eff}(t)=\log \left[G(\vec{0},t)/G(\vec{0},
  t+1)\right]\to E_0$. Utilizing a basis of interpolating fields
$J_N(x)$ one can extract both the ground state and excited states.  In
Fig.~\ref{fig:spectrum} we show an example of utilizing a
five-dimensional basis of interpolators to extract the excited states
of the nucleon in the positive parity channel. As can be seen, identifying a
plateau region is crucial to reliably extract the energy of excited
states. Using more suitable interpolating fields can improve the
plateaus as shown in Fig.~\ref{fig:spectrum2} for both the positive
and negative parity channels.

\begin{figure}
  \centering
  \begin{minipage}{0.475\linewidth}
    \includegraphics[width=\linewidth]{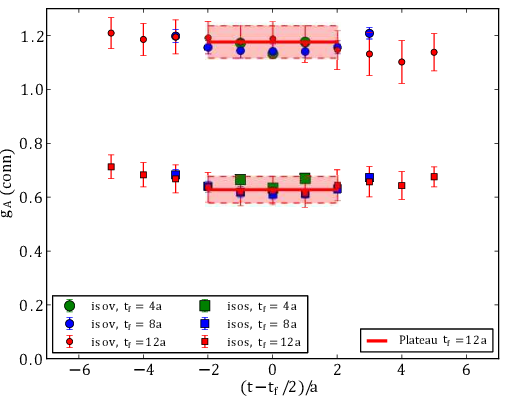}
    \caption{The nucleon matrix element of the  isovector (isov) and the connected contribution to
      the isoscalar (isos)  axial-vector operator. The solid
      line shows the fit to a constant.}\label{fig:plateau_conn}
  \end{minipage}
  \hfill
  \begin{minipage}{0.475\linewidth}\vspace*{-0.5cm}
    \includegraphics[width=\linewidth]{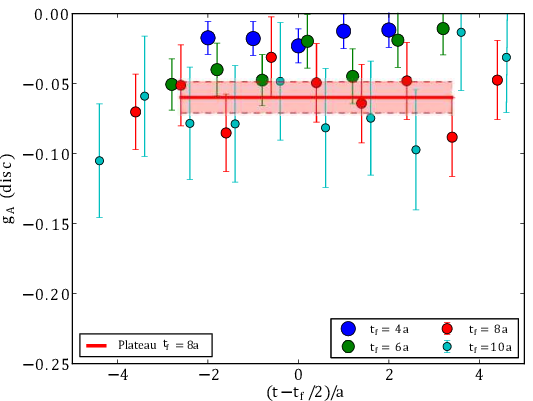}
    \caption{The disconnected contribution to the nucleon matrix element of the isoscalar axial operator as a function of the insertion
      time-slice. The solid line shows the fit
      to a constant.}\label{fig:plateau_disc}
  \end{minipage}
\end{figure}

Unlike the case of the isovector nucleon matrix elements, where quark disconnected contributions cancel, for the isoscalar case,  both
connected and disconnected quark diagrams contribute. These contributions
are shown in
Fig.~\ref{fig:plateau_conn}, for the case of the isoscalar nucleon matrix 
element of the  axial-vector current at $q^2=0$.  
Results for the ratio of Eq.~(\ref{ratio}) are shown  for various source-sink
time separations. When the ground state dominates there is convergence
of the results obtained for various source-sink time separations. 

The disconnected diagrams are notoriously difficult to compute because
they exhibit large statistical errors and require knowledge of all
elements of the quark propagator. The so called all-to-all propagator
is impractical to compute and store explicitly. Stochastic techniques
are typically employed to obtain estimates of the all-to-all
propagator combined with techniques to efficiently multiply the number
of measurements to suppress the statistical uncertainty. Such
calculations have been shown to be especially suited for
GPUs~\cite{Alexandrou:2012py, Alexandrou:2012zz}. In
Fig.~\ref{fig:plateau_disc} we show the disconnected contribution to
the isoscalar nucleon matrix element of the axial-vector current for
various source-sink separations. Note the much larger errors as
compared to those obtained from the connected part.

\vspace{-1ex}

\begin{figure}
  \centering
  \begin{minipage}{0.475\linewidth}
    \includegraphics[width=\linewidth]{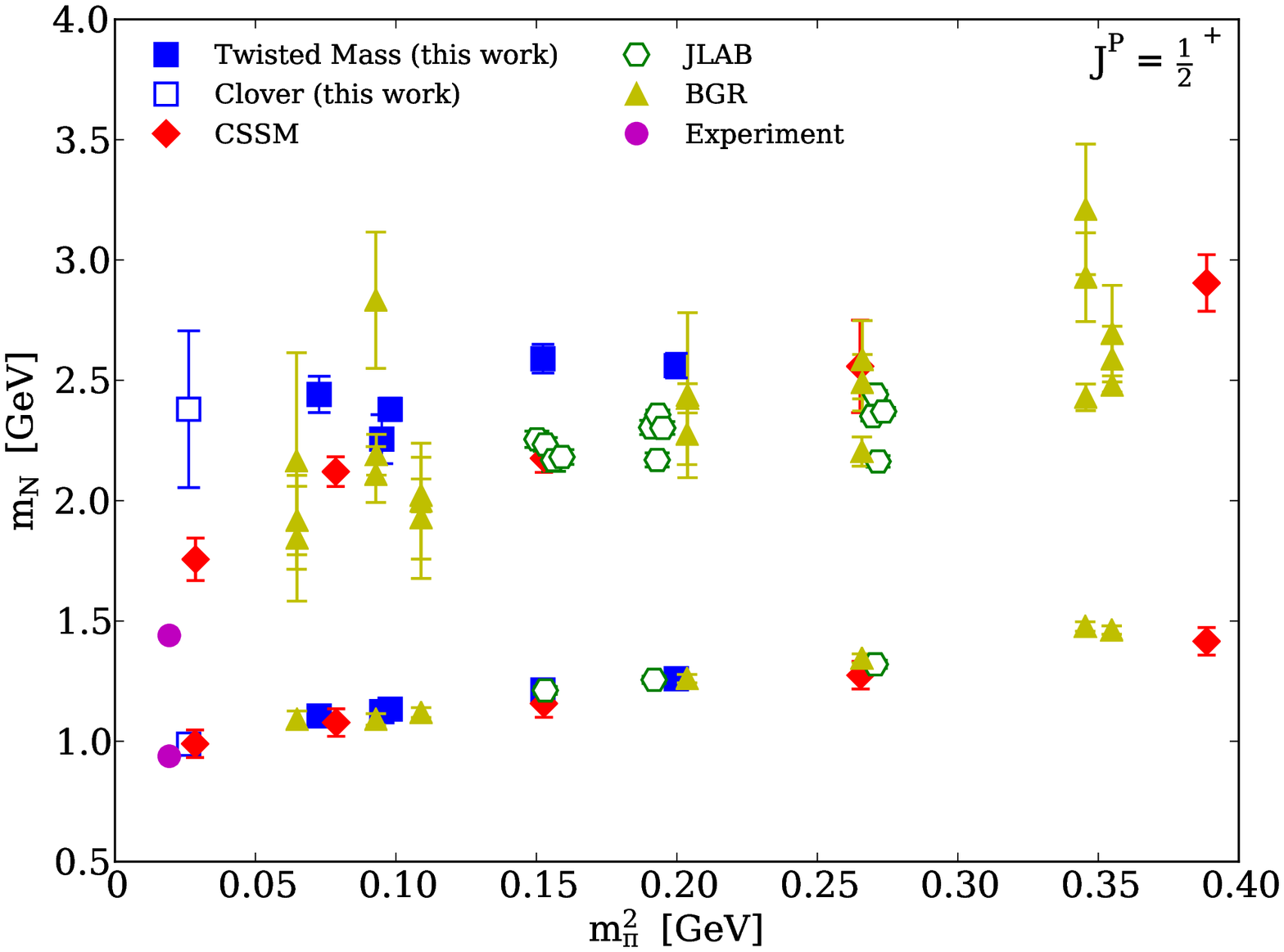}
    \caption{The positive parity states of the nucleon using twisted
      mass and clover fermions from various groups. The magenta
      filled circles show the experimental values of the nucleon
      mass and the Roper.}\label{fig:pos_par}
  \end{minipage}\hfill
  \begin{minipage}{0.475\linewidth}
    \includegraphics[width=\linewidth]{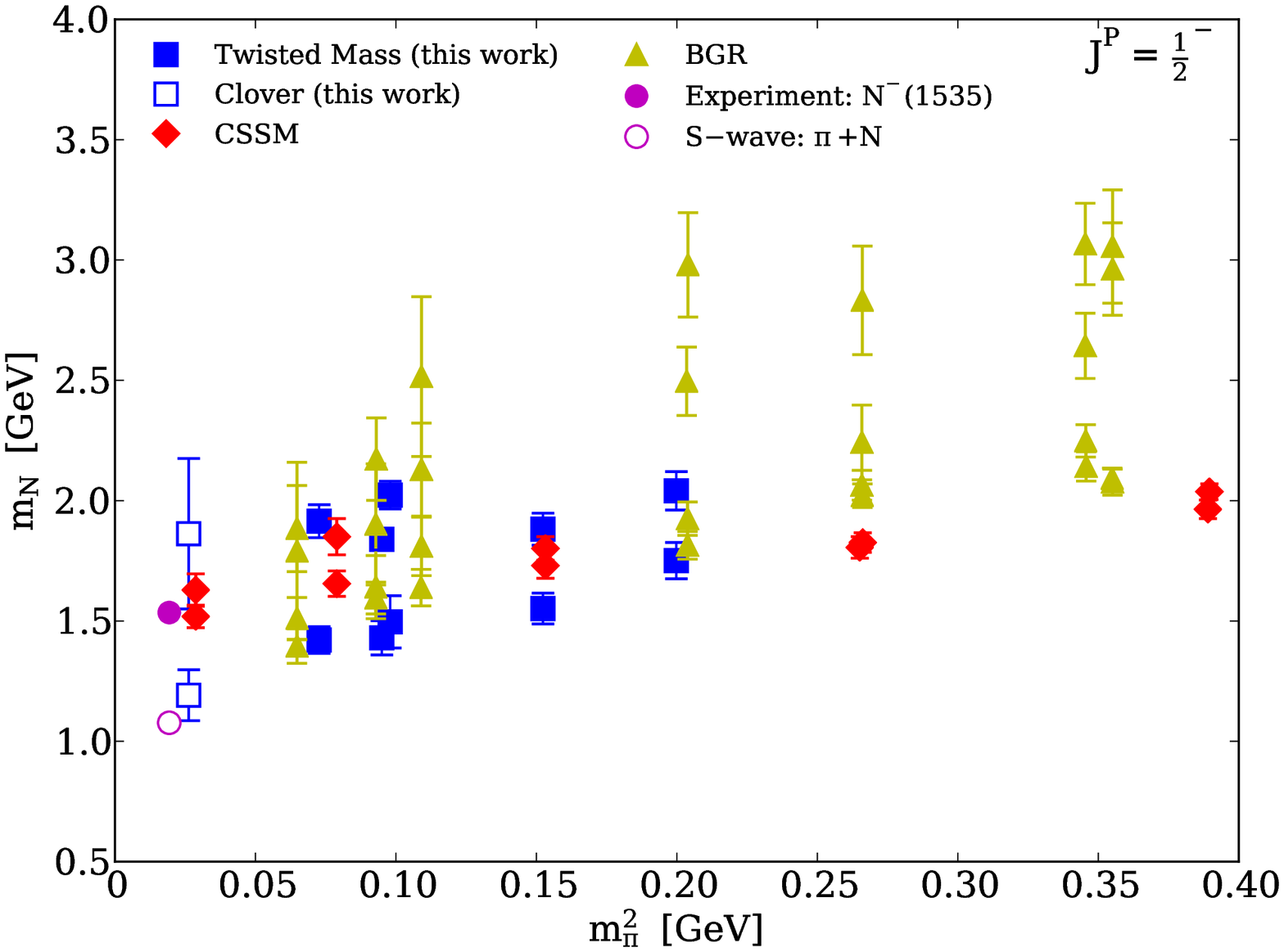}
    \caption{The negative parity states of the nucleon using twisted
      mass and clover fermions from various groups. The filled
      magenta circle shows the $S_{11}(1535)$, whereas the open
      magenta circle the $\pi N$ mass.}
    \label{fig:neg_par}
    \end{minipage}
\end{figure}

\section{Results}
The spectrum of the nucleon is explored by several groups. In
Figs.~\ref{fig:pos_par} and \ref{fig:neg_par} we show recent results
on the first excited states of the
nucleon~\cite{Alexandrou:2013fsu}. All lattice results are in nice
agreement for the ground state. Results for the first
excited state show larger statistical errors and deviations. Most
lattice data are higher than the mass of the Roper, even when pion
masses at almost the physical value are used. This is currently
under investigation. In the negative parity channel, lattice results give the
experimental value of the $S_{11}(1535)$.

\begin{figure}
  \centering
    \begin{minipage}{0.475\linewidth}
      \includegraphics[width=\linewidth]{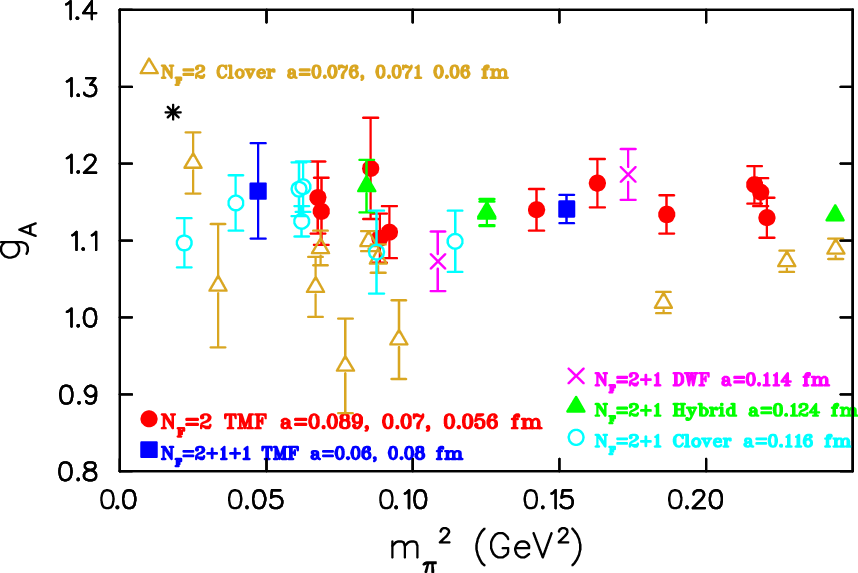}
      \caption{Lattice results on the axial charge of the
        nucleon for: $N_f{=}2$ and $N_f=2+1+1$
        TMF~\protect\cite{Alexandrou:2013joa}; $N_f=2+1$
        DWF~\protect\cite{Yamazaki:2009zq};
        hybrid~\protect\cite{Bratt:2010jn}; $N_f{=}2$
        clover~\protect\cite{Horsley:2013ayv}; $N_f{=}2+1$
        clover~\protect\cite{Green:2012rr}. The physical point is
        shown by the asterisk.}
      \label{fig:gA}
    \end{minipage}\hfill
    \begin{minipage}{0.475\linewidth}
      \includegraphics[width=\linewidth]{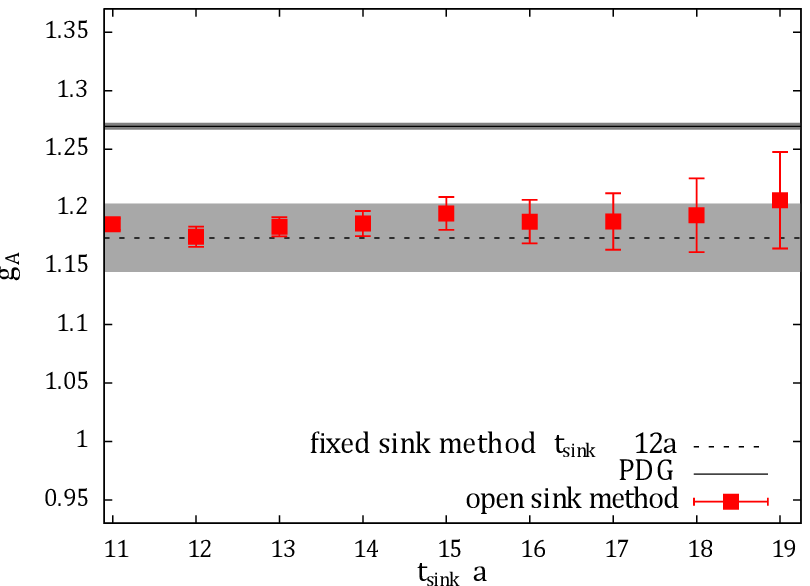}
      \caption{The value of $g_A$ versus the
        sink-source time separation  ($N_f=2+1+$ TMF at  $\sim$350~MeV.}\label{fig:gA excited}
    \end{minipage}
\end{figure}

Reproducing the nucleon axial charge is important for benchmarking our
lattice QCD formulation beyond reproducing masses of low-lying
hadrons.  We show in Fig.~\ref{fig:gA}, the axial charge of the
nucleon $g_A$, extracted from fitting to the plateau region of the ratio of Eq.~(\ref{ratio}) and  which uses the isovector current for which
disconnected contributions cancel.
 The first observation is the
consistency between different lattice formulations. This consistency
is non-trivial, since the results shown are at different, non-zero
lattice spacings, and in general these different formulations have
different finite-spacing correction terms. Furthermore, all results
underestimate the experimental value as the pion mass approaches its
physical value. Candidates for this discrepancy are currently thought
to be systematic effects in lattice calculations, such as artifacts
caused by the finite volume or contamination of the matrix element by
excited states. The latter has been investigated thoroughly in a high
statistics calculation~\cite{Dinter:2011sg}. As can be seen in
Fig.~\ref{fig:gA excited}, $g_A$ does not suffer from excited state
contamination since the plateau value remains consistent for a large
range of source-sink separations. This conclusion is consistent with
Fig.~\ref{fig:plateau_conn}. An alternative to fitting  the plateau of Eq.~(\ref{ratio}) is the summation method that modifies the  excited state contributions~\cite{Capitani:2012gj,Green:2012ud}. Further sources of systematic
uncertainties are currently under investigation by various lattice QCD
collaborations.  

\begin{figure}
  \begin{minipage}[t]{0.48\linewidth}
    \includegraphics[width=\linewidth]{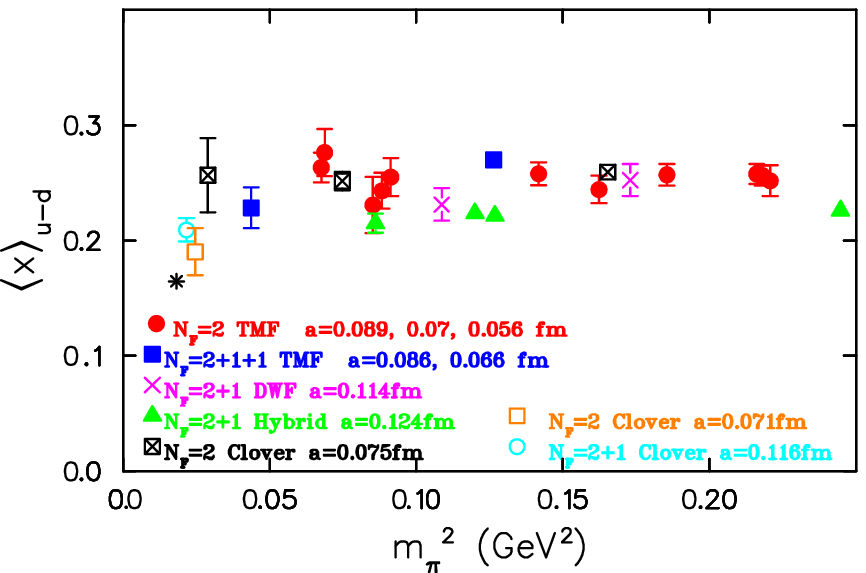}
    \caption{Lattice results on the isovector momentum fraction
      $\langle x \rangle _{u-d}$ of the nucleon. The crosses, crossed 
and open squares are
      from Refs.~\protect\cite{Aoki:2010xg, Pleiter:2011gw,Bali:2012av}, respectively. The rest of the notation is the
      same as that in Fig.~\ref{fig:gA}.}
    \label{fig:ave_x}
  \end{minipage}\hfill
  \begin{minipage}[t]{0.48\linewidth}
    \includegraphics[width=\linewidth]{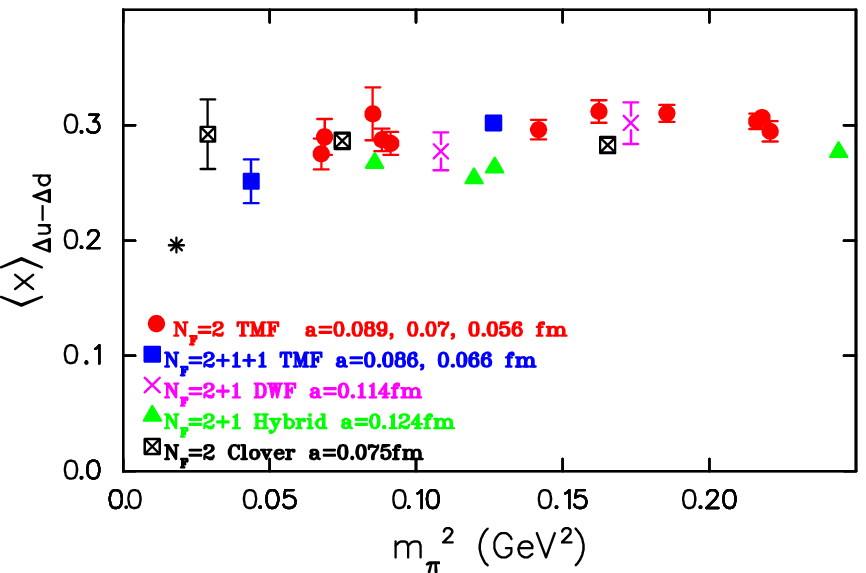}     
    \caption{Lattice results on the isovector polarized moment $
      \langle x \rangle _{\Delta u-\Delta d}$ of the nucleon. The
      notation is the same as that in Fig.~\ref{fig:ave_x}.}
    \label{fig:helicity}    
  \end{minipage}
\end{figure}

In Fig.~\ref{fig:ave_x} we show the isovector combination for the
momentum fraction $A_{20}(0) = \langle x \rangle_{u-d}$ for various
fermion discretization schemes. The general conclusion is again that
lattice results are in agreement among themselves, a strong indication that finite
lattice spacing corrections are smaller than the statistical
errors. Results at almost physical pion mass in this case appear to be
converging to the physical point. Confirmation of this trend is
on-going with computations being performed at near physical pion mass
and large volumes. In Fig.~\ref{fig:helicity} we show results for the
helicity $\tilde{A}_{20}(0) = \langle x \rangle_{\Delta u- \Delta d}$.
The situation is very similar to the momentum fraction.

The matrix elements presented so far are directly related to the
contribution of the nucleon constituent quarks to its spin, as
explained in the introduction. Knowledge of both isovector and
isoscalar combinations are required in order to obtain the individual
quark contributions. From here on, we ignore disconnected
contributions to the isoscalar matrix elements. There are indications
that their contributions are small, e.g. for the case of $g_A =
\tilde{A}_{10}(0)$ these are expected to be of the order of 10\%, as
can be seen from Fig.~\ref{fig:plateau_disc}. Calculations are
underway for their precise determination. The total angular momentum
carried by quarks requires knowledge of $B_{20}(0)$, which on the
lattice cannot be obtained at zero momentum transfer. $B_{20}(0)$ is
therefore calculated by fitting the momentum dependence, for which
 we assume a linear form.

\begin{figure}
  \centering
  \includegraphics[width=\linewidth]{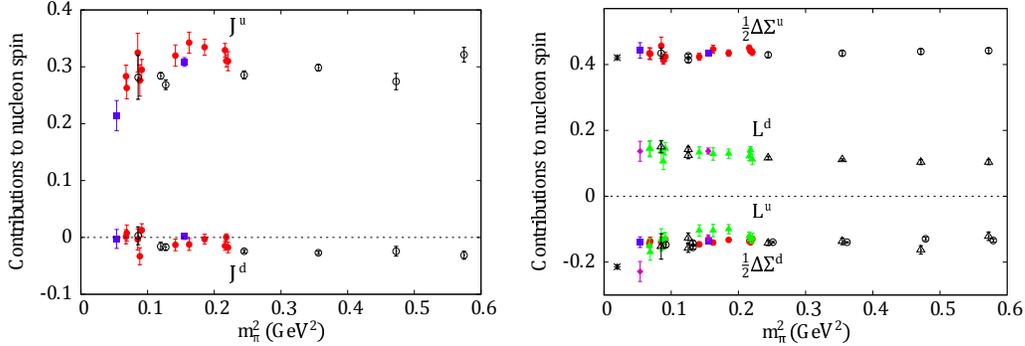}     
  \caption{Quark contributions to the proton spin as a function of
    the pion mass. Left: the total contributions of the up- and
    down-quarks ($J^u$ and $J^d$ respectively). Filled squares are
    obtained using $N_f=2+1+1$ TMF, filled circles
    using $N_f=2$ TMF, compared to results obtained using a
    hybrid action~\protect\cite{Bratt:2010jn} (open symbols). Right:
    The quark spin and quark orbital angular momentum contribution to
    the nucleon spin obtained using $N_f=2+1+1$ twisted mass fermions
    (filled squares and diamonds respectively), using $N_f=2$ twisted
    mass fermions (filled circles and triangles respectively) and
    using a hybrid action (open circles and open triangles
    respectively). Asterisks denote the experimental results.}
  \label{fig:spin_others}\vspace*{-0.3cm}
\end{figure}

We show in Fig.~\ref{fig:spin_others} the quark contributions to the
proton spin. We compare results obtained using $N_f=2$ and $N_f=2+1+1$
TMF with those using a hybrid action of $N_f=2+1$
staggered sea and domain wall valence quarks~\cite{Bratt:2010jn}.  As
can be seen, the total spin carried by the u- and d-quarks $J^{u+d}
\sim J^u$ since $J^d \sim 0$. This can be traced to the intrinsic spin
$\frac{1}{2}\Delta\Sigma^d$ and angular momentum $L^d$ of the d-quarks
canceling.

\begin{figure}
\centering
  \begin{minipage}{0.475\linewidth}
    \includegraphics[width=\linewidth]{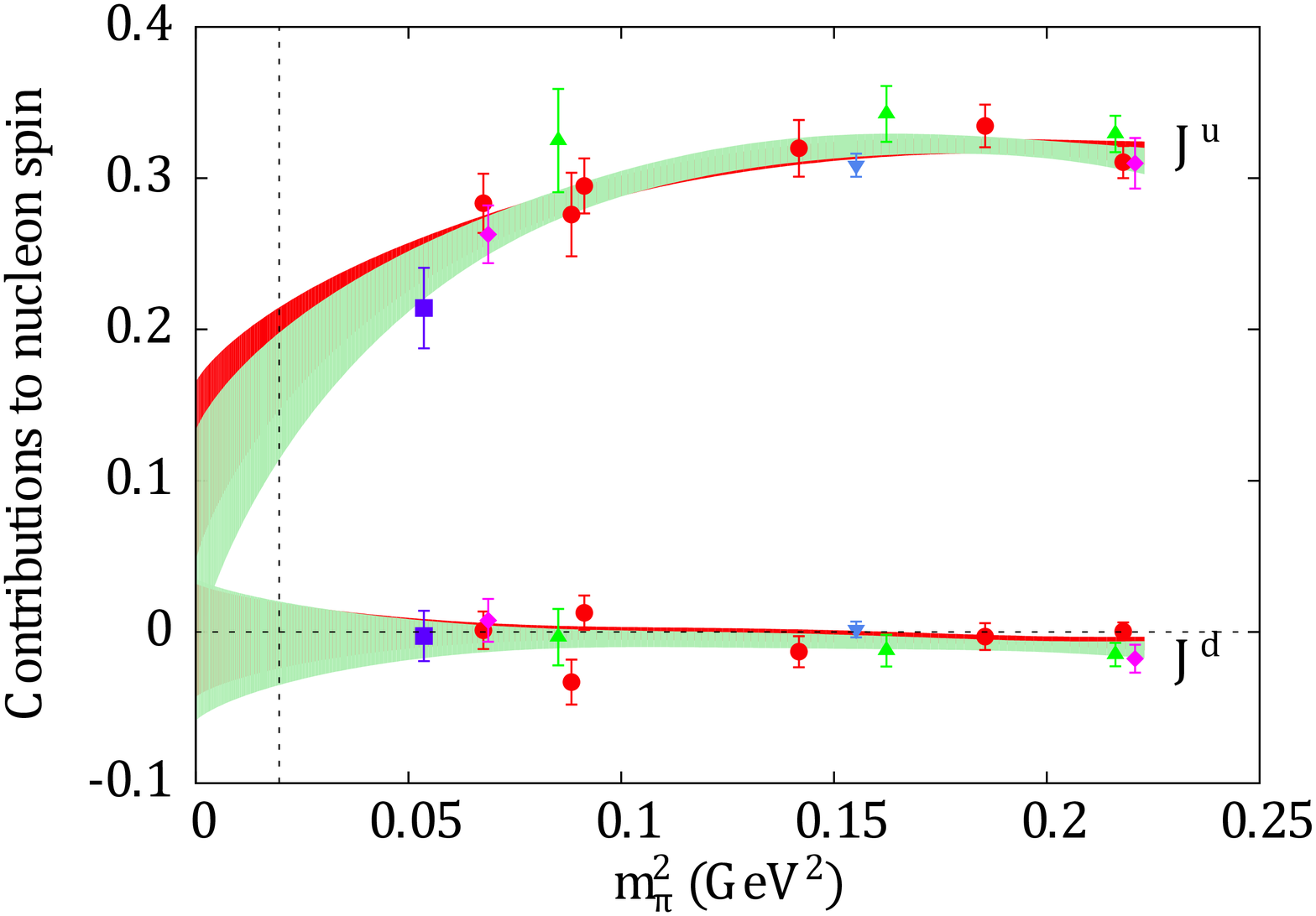}
  \end{minipage} \hfill    
  \begin{minipage}{0.475\linewidth}
    \includegraphics[width=\linewidth]{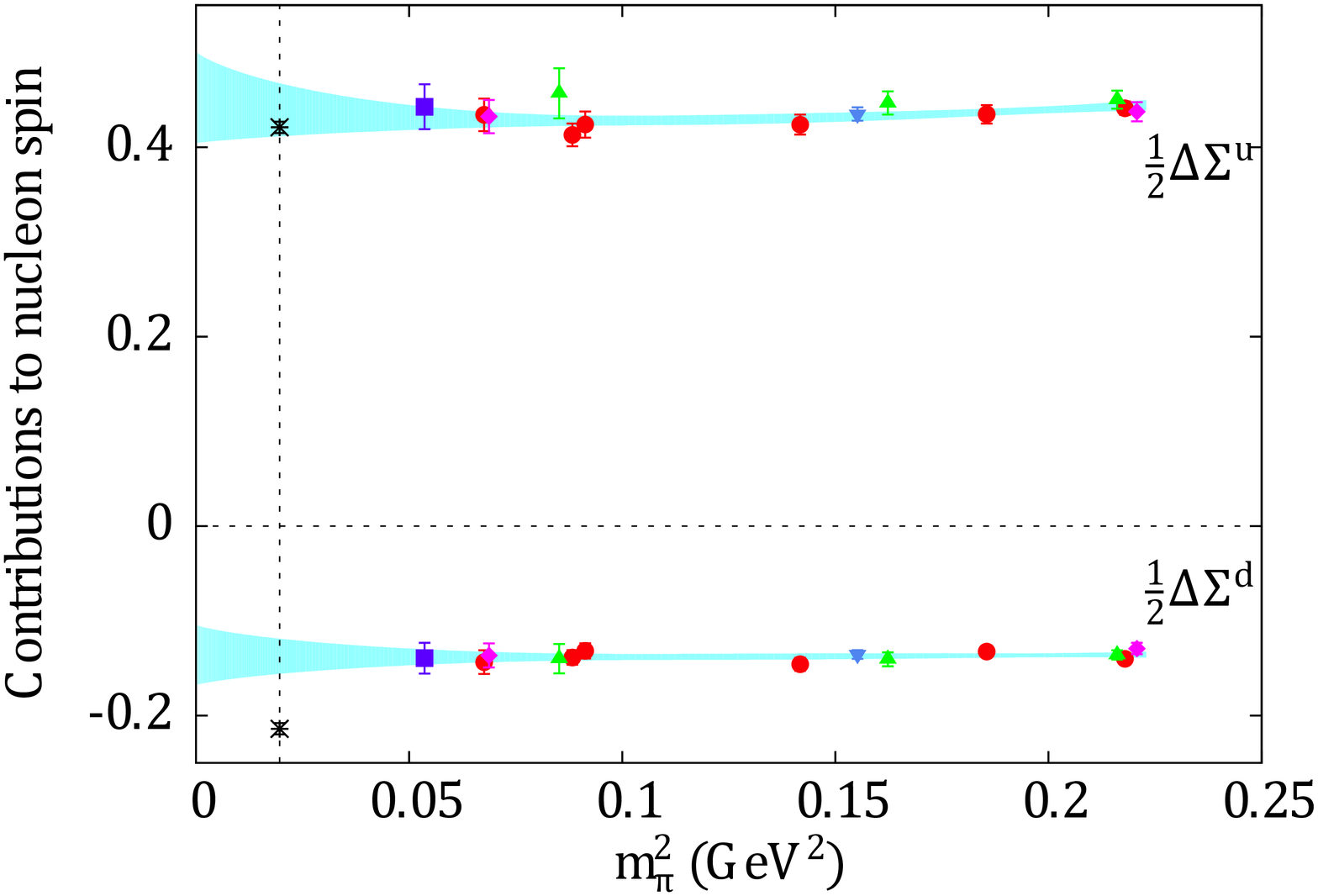}
  \end{minipage}     
  \caption{Chiral extrapolation of the quark contributions to the
    proton spin. The vertical line denotes the physical point. The
    data are obtained using $N_f=2$ twisted mass with $a=0.089$,
    $0.07$ and $0.056$~fm (red circles, green triangles and magenta
    diamonds), and $N_f=2+1+1$ twisted mass with $a=0.086$ and
    $0.066$~fm (blue inverted triangles and blue squares).}
  \label{fig:spin_extrap}
\end{figure}

To obtain some insight on the spin carried by the quarks at the
physical point we use Heavy Baryon Chiral Perturbation Theory
(HB$\chi$PT) to extrapolate the lattice data. The expressions for
these fits are given in Ref.~\cite{Alexandrou:2013joa}, where also covariant
$\chi$PT was used giving qualitatively similar results. A source of
systematic error may came from the fact that $B_{20}(0)$ is obtained
via a fit. This is done by extrapolating the data obtained by using
two fit ranges in the fit to $B_{20}(Q^2)$, shown by the red and green
bands in Fig.~\ref{fig:spin_extrap}. As can be seen both fit ranges
yield consistent results. For the case of the intrinsic spin where
experimental data are available, a comparison shows that the
down-quark contribution is under estimated, while for the u-quark
there is agreement. However, this disagreement is too small to account
for the missing 50\% of the spin of the nucleon. Disconnected
contributions to the quark spins are also expected to be small and
therefore a large fraction of the proton spin is not carried by the
quarks.

\section{Conclusions}
A review of recent lattice QCD nucleon structure calculations is
presented, comparing the results obtained by different groups.  The
nucleon first excited states are presented with good agreement with
experiment in the negative parity channel but still not being
conclusive in the case of the Roper. Lattice QCD data on the nucleon
axial charge are in agreement among themselves but, however, underestimate the
experimental value for $g_A$ believed to be due to systematic effects appearing in lattice computations. This calls for an in depth study to identify possible
sources of this discrepancy, something which is being investigated
currently by several lattice groups. Lattice data for the momentum
fraction and helicity on the other hand are seen to converge towards
the experimental value as the pion mass decreases below 250~MeV
although still remain higher.  Recent results near the physical point
have suggested excited states~\cite{Green:2012ud} to be the source for
the remaining discrepancy. This is being investigated.

The axial charge, combined with the momentum fraction, give insight on
the fraction of the nucleon spin carried by quarks. Using both
isovector and isoscalar combinations of these quantities, with a
cautionary remark on the omission of disconnected diagrams that
contribute to the latter, we extract the individual up- and down-quark
contributions to the proton spin. Qualitatively, we find agreement
with what is observed experimentally for the quark intrinsic spin
fractions. Recent lattice calculations also agree in indicating that
the total spin fraction carried by down-quarks is much smaller and
near-zero compared to the up-quark and that the quarks account for
about half of the proton spin.

\acknowledgments 
We thank the SFB/TRR-55 for making available to us their gauge
configurations for $\beta=5.29$ and $\kappa=0.1364$ ($m_\pi=160$~MeV)
via the ILDG. We have used HPC resources on the Jugene system at the
research center in J\"ulich through PRACE and the Cy-Tera facility of the Cyprus Institute under
the project Cy-Tera (NEA Y$\Pi$O$\Delta$OMH/$\Sigma$TPATH/ 0308/31),
first access call (project lspro113s1). 
 This work is supported in part
by the Cyprus Research Promotion Foundation under contracts
TECHNOLOGY/$\Theta$E$\Pi$I$\Sigma$/0311 (BE)/16, $\Pi$PO$\Sigma$E$\Lambda$KY$\Sigma$H/NEO$\Sigma$/0609/16, and the Research
Executive Agency of the European Union under Grant Agreement number
PITN-GA-2009-238353 (ITN STRONGnet).  K. J. was supported in part by
the Cyprus Research Promotion Foundation under contract
$\Pi$PO$\Sigma$E$\Lambda$KY$\Sigma$H/EM$\Pi$EIPO$\Sigma$/0311/16. K.J. and V.D. acknowledge financial support by the DFG-funded corroborative research center SFB/TR9.

\vspace*{-0.3cm}

\bibliographystyle{JHEP}
\bibliography{Alexandrou_QCDN12}

\providecommand{\href}[2]{#2}\begingroup\raggedright\begin{mcbibliography}{10}

\bibitem{Alexandrou:2010cm}
\sc C.~Alexandrou {\em PoS} {\bf LATTICE2010} (2010) 001,
  [\href{http://xxx.lanl.gov/abs/1011.3660}{{\tt arXiv:1011.3660}}]
\bibitem{Alexandrou:2011iu}
\sc C.~Alexandrou {\em Prog.Part.Nucl.Phys.} {\bf 67} (2012) 101--116,
  [\href{http://xxx.lanl.gov/abs/1111.5960}{{\tt arXiv:1111.5960}}]
\bibitem{Alexandrou:2012da}
\sc C.~Alexandrou, \sc C.~Papanicolas, and \sc M.~Vanderhaeghen
  \href{http://xxx.lanl.gov/abs/1201.4511}{{\tt arXiv:1201.4511}}
\bibitem{Alexandrou:2012py}
\sc C.~Alexandrou and \sc others {\em PoS} {\bf LATTICE2012} (2012) 184,
  [\href{http://xxx.lanl.gov/abs/1211.0126}{{\tt arXiv:1211.0126}}]
\bibitem{Alexandrou:2012zz}
\sc C.~Alexandrou, \sc K.~Hadjiyiannakou, \sc G.~Koutsou, \sc A.~O'Cais, and
  \sc A.~Strelchenko {\em Comput.Phys.Commun.} {\bf 183} (2012) 1215--1224,
  [\href{http://xxx.lanl.gov/abs/1108.2473}{{\tt arXiv:1108.2473}}]
\bibitem{Alexandrou:2013fsu}
\sc C.~Alexandrou, \sc T.~Korzec, \sc G.~Koutsou, and \sc T.~Leontiou
  \href{http://xxx.lanl.gov/abs/1302.4410}{{\tt arXiv:1302.4410}}
\bibitem{Dinter:2011sg}
\sc S.~Dinter and \sc others {\em Phys.Lett.} {\bf B704} (2011) 89--93,
  [\href{http://xxx.lanl.gov/abs/1108.1076}{{\tt arXiv:1108.1076}}]
\bibitem{Capitani:2012gj}
\sc S.~Capitani and \sc others {\em Phys.Rev.} {\bf D86} (2012) 074502,
  [\href{http://xxx.lanl.gov/abs/1205.0180}{{\tt arXiv:1205.0180}}]
\bibitem{Green:2012ud}
\sc J.~Green and \sc others \href{http://xxx.lanl.gov/abs/1209.1687}{{\tt
  arXiv:1209.1687}}
\bibitem{Alexandrou:2013joa}
\sc C.~Alexandrou and \sc others \href{http://xxx.lanl.gov/abs/1303.5979}{{\tt
  arXiv:1303.5979}}
\bibitem{Yamazaki:2009zq}
\sc T.~Yamazaki and \sc others {\em Phys. Rev.} {\bf D79} (2009) 114505,
  [\href{http://xxx.lanl.gov/abs/0904.2039}{{\tt arXiv:0904.2039}}]
\bibitem{Bratt:2010jn}
\sc J.~D. Bratt and \sc others {\em Phys. Rev.} {\bf D82} (2010) 094502,
  [\href{http://xxx.lanl.gov/abs/1001.3620}{{\tt arXiv:1001.3620}}]
\bibitem{Horsley:2013ayv}
\sc R.~Horsley and \sc others \href{http://xxx.lanl.gov/abs/1302.2233}{{\tt
  arXiv:1302.2233}}
\bibitem{Green:2012rr}
\sc J.~Green and \sc others {\em PoS} {\bf LATTICE2012} (2012) 170,
  [\href{http://xxx.lanl.gov/abs/1211.0253}{{\tt arXiv:1211.0253}}]
\bibitem{Aoki:2010xg}
\sc Y.~Aoki and \sc others {\em Phys.Rev.} {\bf D82} (2010) 014501,
  [\href{http://xxx.lanl.gov/abs/1003.3387}{{\tt arXiv:1003.3387}}]
\bibitem{Pleiter:2011gw}
\sc D.~Pleiter and \sc others {\em PoS} {\bf LATTICE2010} (2010) 153,
  [\href{http://xxx.lanl.gov/abs/1101.2326}{{\tt arXiv:1101.2326}}]
\bibitem{Bali:2012av}
\sc G.~S. Bali and \sc others {\em Phys.Rev.} {\bf D86} (2012) 054504,
  [\href{http://xxx.lanl.gov/abs/1207.1110}{{\tt arXiv:1207.1110}}]
\end{mcbibliography}\endgroup

\end{document}